# Accelerator system for the PRISM based muon to electron conversion experiment

A. Alekou$^{ep}$, R. Appleby $^{ab}$, M. Aslaninejad $^d$, R. J. Barlow $^m$, R. Chudzinski$^q$
K. M. Hock$^{ac}$, J. Garland $^{ab}$, L. J. Jenner $^e$, D. J. Kelliher $^f$, Y. Kuno $^g$, A. Kurup $^d$,
J-B. Lagrange $^h$, M. Lancaster $^i$, S. Machida $^f$, Y. Mori $^h$,  B. Muratori $^{aj}$,
C. Ohmori $^k$, H. Owen $^{ab}$, J. Pasternak $^{d*}$, T Planche $^n$, C. Prior $^f$,  A. Sato $^g$, Y. Shi$^q$,
S. Smith $^{aj}$,  Y. Uchida $^d$, H. Witte$^o$  and T. Yokoi $^l$

$^a$*Cockcroft Institute, Warrington, UK*
$^b$*University of Manchester, UK*
$^c$*University of Liverpool, UK*
$^d$*Imperial College London, UK*
$^e$*previously at Imperial College London, UK*
$^f$*STFC-RAL-ASTeC, Harwell, UK*
$^g$*Osaka University, Osaka, Japan*
$^h$*Kyoto University, KURRI, Osaka, Japan*
$^i$*UCL, London, UK*
$^j$*STFC-DL-ASTeC, Warrington, UK*
$^k$*KEK/JAEA, Ibaraki-ken, Japan*
$^l$*previously at JAI, Oxford University , UK*
$^m$*University of Huddersfield, UK*
$^n$*TRIUMF, Canada*
$^o$*BNL, USA*
$^p$*CERN, Geneva, Switzerland*
$^q$*UROP Student Programme, Imperial College London, UK*

*E-mail:* j.pasternak@imperial.ac.uk

ABSTRACT: The next generation of lepton flavour violation experiments need high intensity and high quality muon beams. Production of such beams requires sending a short, high intensity proton pulse to the pion production target, capturing pions and collecting the resulting muons in the large acceptance transport system. The substantial increase of beam quality can be obtained by applying  the RF phase rotation on the muon beam in the dedicated FFAG ring, which was proposed for the PRISM project. This allows to reduce the momentum spread of the beam and to purify from the unwanted components like pions or secondary protons. A PRISM Task Force is addressing the accelerator and detector issues that need to be solved in order to realize the PRISM experiment. The parameters of the required proton beam, the principles of the PRISM experiment and the baseline FFAG design are introduced. The spectrum of alternative designs for the PRISM FFAG ring is shown. The ring injection/extraction system, matching with the solenoid channel and progress on the ring's main hardware systems like RF and kicker magnet are presented. The current status of the study and its future directions are discussed.

---

$^*$Corresponding author.

# 1. Introduction

The results from the numerous neutrino oscillation experiments showed the existence of the nonzero neutrino mass and the violation of the lepton flavour conservation law for the neutral leptons. This experimental evidence clearly shows that the current theory of particle physics – the Standard Model (SM) is incomplete and requires modifications. The search for physics beyond the SM will be continued especially with the Large Hadron Collider (LHC) at the energy frontier, but also with neutrino experiments and other precision measurements, which coupled with the future high intensity proton accelerators like the proposed Project-X at Fermilab, will substantially increase their physics potentials. The non-conservation of the lepton number in the neutrino sector puts a question upon existence of charged lepton flavour violation processes, such as muon to electron conversion. This process promises to be a fruitful area to search for physics beyond the SM, as its existence would put very strong constraints on possible new theoretical models. The COMET [1] and Mu2e [2] experiments have been proposed to measure muon to electron conversion with a sensitivity to the branching ratio smaller than $10^{-16}$. In order to achieve a better sensitivity of $<10^{-18}$, the PRISM (Phase Rotated Intense Slow Muon beam) system was proposed. This unprecedented sensitivity can be realised by using an FFAG (Fixed Field Alternating Gradient) ring, which will allow for reduction of muon beam energy spread by longitudinal phase-space rotation using RF system and simultaneous beam purification from unwanted contaminations like pions and secondary protons. This will allow the creation of a high quality muon beam with an excellent experimental capabilities. This proposal has become realistic after substantial progress in the field of FFAG accelerators in recent years. However, there is still a number of technological challenges that need to be addressed before a design for the experiment can be realized. The PRISM task force was set up to address these issues and utilise synergies with other projects such as the Neutrino Factory [3] and the Muon Collider and more recently also nuSTORM[4].

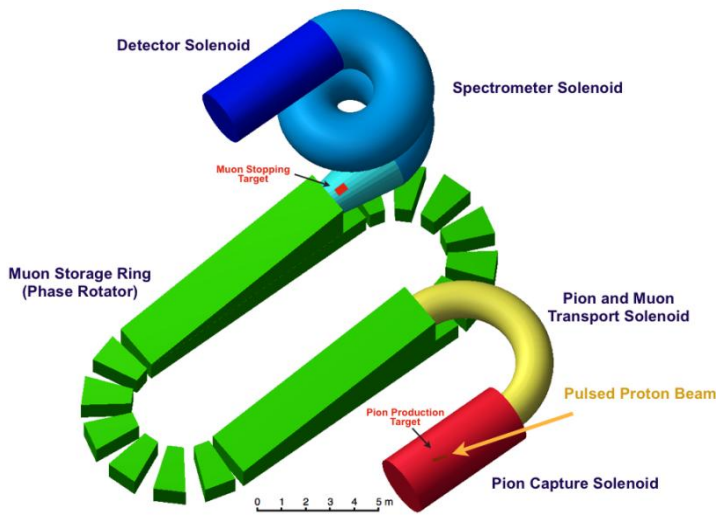

Figure1: Conceptual layout of the PRISM accelerator and experimental system.

The conceptual schematic of the PRISM accelerator and experimental system is shown in Figure 1. The PRISM requires that a high intensity short duration proton bunch is sent to the target immersed in high field solenoid, where pions are produced and captured backwards, which allows to achieve a good low energy pion yield simultaneously reducing the high energy background and simplifying the primary proton dump issues. The pion beam is then transported in the series of bent solenoids decaying into muons. Muon beam with still large pion contamination and huge energy spread is then injected into an FFAG ring where in a few turns RF phase rotation is performed reducing the energy spread by an order of magnitude and allowing to the unwanted pions to decay. Following extraction from the FFAG beam is sent to the muon stopping target, where muons are allowed to be stopped and captured by nuclei. As the signal of muon to electron conversion is a mono-energetic electron with 105 MeV/c, which is just above the continuous muon decay spectrum, a spectrometer optimised for this momentum can be used before the dedicated PRISM detector system (PRIME).

## 1.1 Status of PRISM accelerator R&D

The PRISM project was proposed in order to realize a low-energy muon beam with a high-intensity, narrow energy spread and high purity. For this purpose, a scaling FFAG ring has been chosen as for this solution a large transverse and longitudinal acceptance is guaranteed. The initial design of the FFAG ring for PRISM is based on a lattice with 10 identical DFD triplets. An intensive R&D program to study its feasibility was performed during 2003 - 2009 in Osaka, Japan. During this program, full size large aperture scaling FFAG magnets and an RF system with Magnetic Alloy (MA) cores were successfully developed with the designed performance. In the next phase the ring accelerator with the six magnets was assembled (as shown in Figure 2) and the successful phase rotation of alpha particles was demonstrated [5].

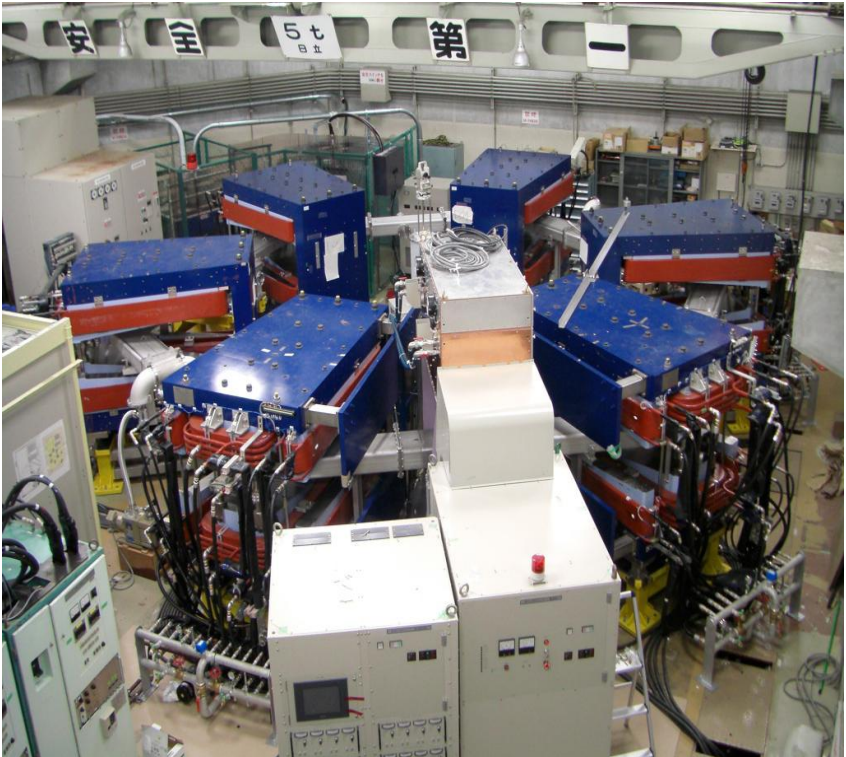

Figure 2: Layout of the 6-cell scaling FFAG ring assembled at RCNP in Osaka to perform phase rotation demonstration experiment using alpha particles.

**1.2 PRISM Task Force Initiative**

In order to continue the research towards the PRISM FFAG addressing the remaining difficulties and to strengthen the activity on muon accelerator physics, the PRISM Task Force was formed. The research programme focuses on the design of the transport line from the solenoidal pion decay channel into the FFAG ring, and also on the injection and extraction systems. The studies covers also the alternative FFAG ring design, which could be superior to the current baseline solution. This search aims on a ring design with a very high transverse acceptance and long straight sections in order to facilitate injection and extraction. Possible solutions include: advanced FFAG rings, which are formed from scaling FFAG arcs matched to FFAG-type straight sections; FFAG rings with superperiodicity; and non-scaling FFAG rings.

**2. Proton driver for PRISM**

The use of the phase rotation in the FFAG ring sets the constraints on the proton bunch length at the pion production target. The muon energy spread after pion decay is very large due to initial pion energy spread at the production and the decay kinematics. The muon energy spread after production is fixed by physics and it needs to be large in order to deliver enough muon intensity. On the other hand the muon bunch length can be controlled by minimising the initial proton bunch length at the target. After injection into the FFAG ring the short muon bunch with a large energy spread will be converted into a long one with a small energy spread in the process of the RF phase rotation. It is estimated that the proton bunch length at the pion production target needs to be equal or smaller than 10 ns in total. Another constraint on the proton driver is set by the energy, which needs to be large enough for the efficient pion production, but below the antiproton production threshold, which could be an additional source of background for the experiment. These requirements set the proton driver energy to be in the range of 2-8 GeV. Such a proton driver could be accommodated in several locations following an upgrade of the existing facilities or construction of new accelerator systems. The potential locations include: JPARC in Japan, Project-X at Fermilab in the USA, ISIS after multi-MW upgrade in the UK. In fact any proton driver constructed for the Neutrino Factory or the Muon Collider in the future would have parameters similar to the one needed for PRISM.

    The use of the CW $H^-$ linac at the Project X [6] would require beam accumulation by stripping charge exchange injection in a dedicated ring. The bunch compression would then be performed in the same ring by a rapid rise of the RF voltage or in another dedicated ring, where beam acceleration could also be performed. It is believed that a repetition rate as high as up to 1 kHz could be obtained. Both Step II and Step III foreseen for the Project X will have adequate energy for the PRISM system.

**3. Injection system for PRISM ring**

**3.1 Design of the injection line**

The solenoidal channel has been identified as the most efficient and cost effective pion decay channel, and it is where the muon beam is formed. Its optical properties are characterised by small and equal betatron functions in both transverse planes. The FFAG ring on the other hand requires different betatron functions and a non-zero dispersion function. It is necessary to perform the matching of beam conditions for the broad momentum range of $\pm\,20\%$.

Bend solenoidal channels in S and C shapes were studied using G4Beamline code. For the PRISM application S-channel seems more suitable as the final dispersion function is smaller especially in the S-channel with anti-symmetric dipole correction. The achieved transmission was also higher in the S-channel. At the end of the S-channel optical condition needs to be adjusted such that the beam can be transported into the Alternating Gradient (AG) channel with minimum losses. The adiabatic section was studied, where the solenoidal field would be slowly decreased and the betatron function simultaneously increased. This section was then followed by the final solenoidal cell. By manipulating the length of the adiabatic switch and the field in the final cell, a good matching to the downstream quad pumplet was achieved.

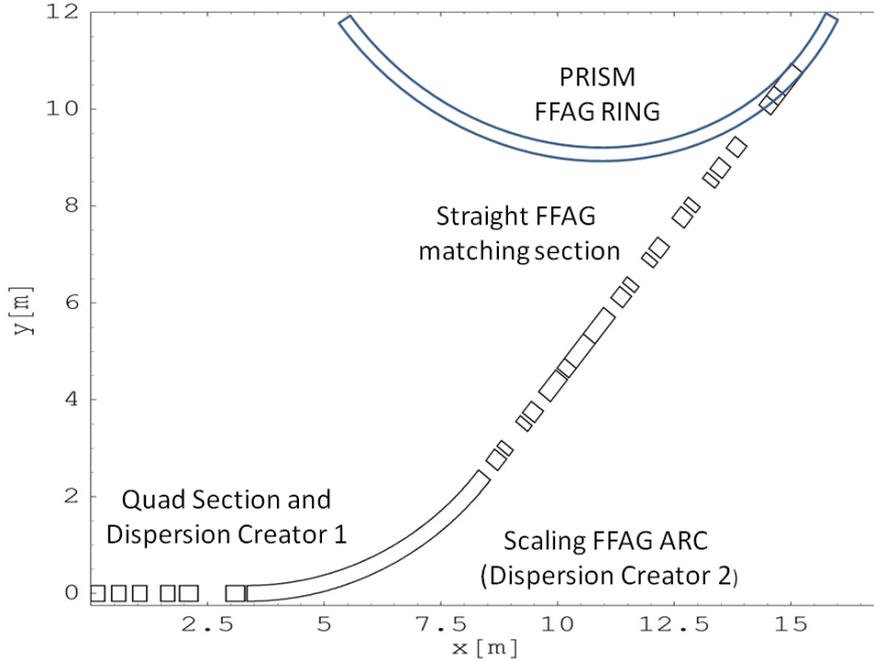

Figure 3: The approximate layout of the AG part of the PRISM muon front end. Dispersion Creator 1 consists of the two rectangular dipoles opposite in bending strength and Dispersion Creator 2 would be made of scaling FFAG cells with $\pi$ horizontal phase advance.

The pumplet (system of 5 quadrupoles) was proposed to act as an intermediate transport channel between the solenoids and the FFAG injection line further downstream. The dispersion function is initially created in the pair of rectangular dipoles identical in size but exactly opposite in the bending strength (Dispersion Creator 1) and further magnified to the final value required by an FFAG dispersion matching section (Dispersion Creator 2). Tracking studies performed till this point showed very good transmission of 97%.

As the injection into the PRISM ring was proposed to be in the vertical direction, the vertical dispersion needs to be matched to zero inside the ring, while the horizontal dispersion must be preserved. This condition must be performed with minimal losses for beam with very large emittances and energy spread. Combination of vertical dipoles and horizontal FFAG magnets are under study to fulfil this challenging goal. The approximate layout of the AG part of the muon front end for the PRISM FFAG ring is shown in Figure 3.

**3.2 Injection scenarios**

The relatively large orbit excursion (~0.46 m) in the reference PRISM FFAG ring design and very large incoming muon beam emittances in both planes, make vertical injection/extraction the only option. The initial ideas for injection/extraction assumed sharing the kicker magnets for both injection and extraction. This solution increases the number of drift spaces, which could be used for the RF cavities, but also the complexity of the kicker magnets. It was proposed to design the separate injection and extraction systems in order to reduce the kicker magnet complexity and further increase the purity of the final beam by decreasing the probability of propagation for the injected particles directly to the extraction line due to the larger distance between the two lines. The injection system will use a vertical septum magnet followed by 2 kicker magnets. Due to the large size of the beam the aperture of the constructed baseline scaling FFAG magnet is not sufficient to transport the full beam between the septum and the first kicker magnet. To solve this problem the design of dedicated insertion cells for injection and extraction with larger aperture magnets are required. Although those insertions will break the symmetry of the ring, the large dynamical acceptance can be preserved, if the phase advances are carefully chosen. The tracking studies using idealized models of the FFAG magnets suggest that this is indeed possible.

### 3.3 Studies of kicker magnet for PRISM

The injection and extraction system for PRISM require kicker magnets with very large apertures due to the large muon beam emittance and orbit excursion. The relatively small ring size and the long muon bunch length after phase rotation dictate a difficult constraint on the rise time of the extraction kicker, which needs to be about 50-60 ns. The initial design of the kicker systems assumes the use of a Pulse Forming Network (PFN) followed by a fast thyratron switch, connected to the kicker magnet by coaxial wires and terminated with a matching resistor.

In order to suppress reflections, the impedance needs to be matched throughout the system. The kicker magnet needs to be subdivided into smaller sub-kickers in order to meet the rise/fall time requirements. Each section of the kicker magnet requires added capacitance in order to match the PFN impedance. In order to obtain a very high repetition rate (~1 kHz) the thermal analysis needs to be performed and a dedicated cooling system may be required. In order to prove the kicker life-time, hardware tests would be needed. The main parameters of the kicker system are summarized in Table 1 and its conceptual schematic is shown in Figure 4.

Table 1. Parameters of the kicker system for the PRISM FFAG ring

| Parameter | Value |
| --- | --- |
| Length/structure | 1.6 m/8 sub-kickers/5 sections each |
| Max B field | 0.02 T |
| Aperture | 0.95 x 0.5 m |
| Flat top (injection/extraction) | (40/210) ns |
| Fall/rise time (injection/extraction) | ~(200/80) ns |
| Stored energy | ~186 J |
| Voltage | 80 kV |
| Impedance | 3 Ohm |
| Max current | 16 kA |
| Inductance | 3 uH |

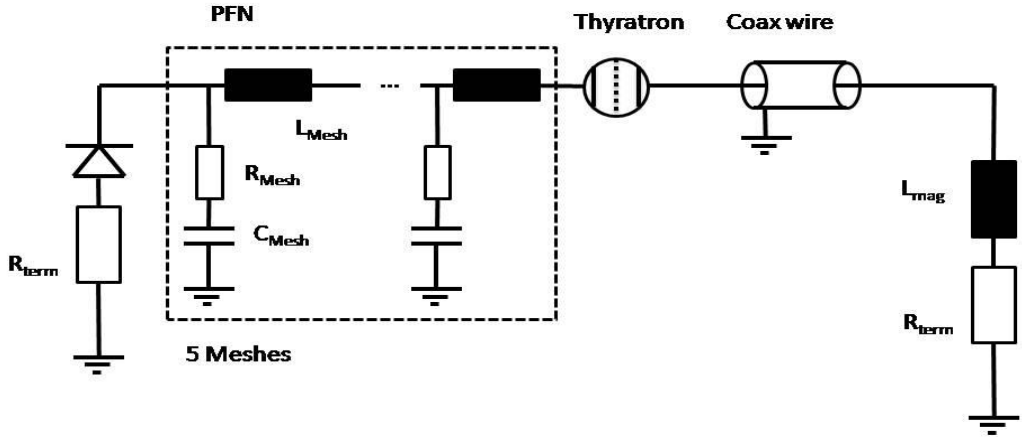

Figure 4: Conceptual schematic of the kicker system for PRISM FFAG ring.

## 4. Alternative ring designs

### 4.1 Advanced FFAG

The problem of injection and extraction remains to be the most important obstacle for realizing the PRISM system. To solve this problem, we consider the use of an advanced FFAG concept [7], in which straight FFAG cells [8, 9] with zero net deflection and magnetic field on the median plane described by $\sim e^{mx}$ are combined with a compact scaling FFAG arc. The layout of the new proposed design is shown in Fig. 5 and the lattice parameters are summarized in Table 2.

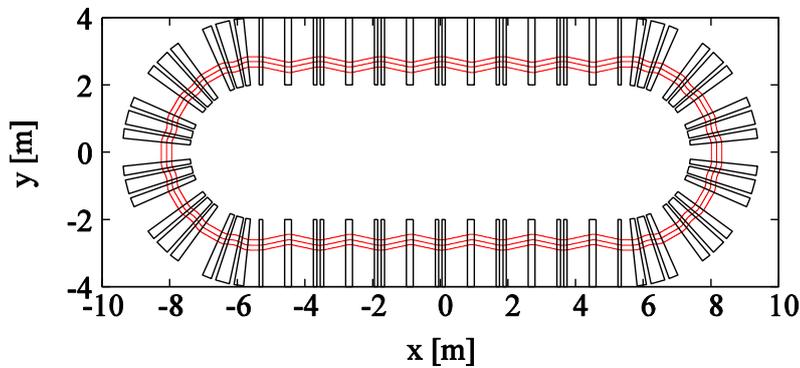

Figure 5: Layout of the advanced FFAG solution for the PRISM ring. Closed orbits of 55 MeV/c, 68 MeV/c and 82 MeV/c muons are shown.

Table 2. Parameters of the advanced FFAG lattice for PRISM

| **Circular section FDF triplet scaling FFAG cell** | |
|---|---|
| k | 2.55 |
| Mean radius (at 68 MeV/c) | 2.7 m |
| Horizontal phase advance | 60 degrees |
| Vertical phase advance | 90 degrees |
| Number of circular cells | 12 |
| **Straight section FDF triplet scaling FFAG cell** | |
| m | 1.3 m$^{-1}$ |
| Cell length | 1.8 m |
| Horizontal phase advance | 27 degrees |
| Vertical phase advance | 97 degrees |
| Number of straight cells | 12 |

Particle tracking studies have been performed using the Runge-Kutta integration in soft edge fields with linear fringe field falloffs. Components of the field off the mid-plane are obtained from a first order Taylor expansion, satisfying Maxwell's equations. The original PRISM design has a very large dispersion function (~1.2 m) that makes the injection and the extraction difficult. The new proposal starts from a smaller one (~0.8 m). A good matching of the periodic beta-functions of the different cells gives a less modulated beta-function, and helps to have a larger acceptance. The first step is thus to minimize the mismatch of the beta-functions, then the bending part of the ring is made transparent by imposing the modulo π phase advance, to limit the effect of the remaining mismatch on the amplitude of the betatron oscillations. The following step is to choose the working point in the tune diagram so that it is far from the structural normal resonances. The transverse acceptance in both planes is studied by tracking over 30 turns a particle with a displacement from the closed orbit and a small deviation in the other transverse direction (~1mm). Collimators (~1m in horizontal direction, ~30 cm in vertical one) are used to identify the lost particles. Horizontal (~24000 π.mm.mrad) and vertical (~6000 π.mm.mrad) acceptances have been estimated from the tracking studies.

An interesting variation of an FFAG ring design for PRISM has been proposed, which is based on advanced scaling FFAGs incorporating arc sections with different radii ("egg-shaped", see Fig. 6). In this design there is more flexibility in performing the matching between the different sections and promising tracking results have been obtained [10].

**4.2 Non-scaling FFAG**

Another approach for having a very large transverse acceptance is to use a non-scaling FFAG. So far only such machine, EMMA (Electron Model with Many Applications) [11] exists. It is presently under commissioning at the Daresbury Laboratory in the UK, which already proved that such machine can indeed be built and operated. Moreover during EMMA commissioning a novel type of charge particle acceleration using RF fields in the so called "serpentine channel" has been demonstrated for the first time [12].

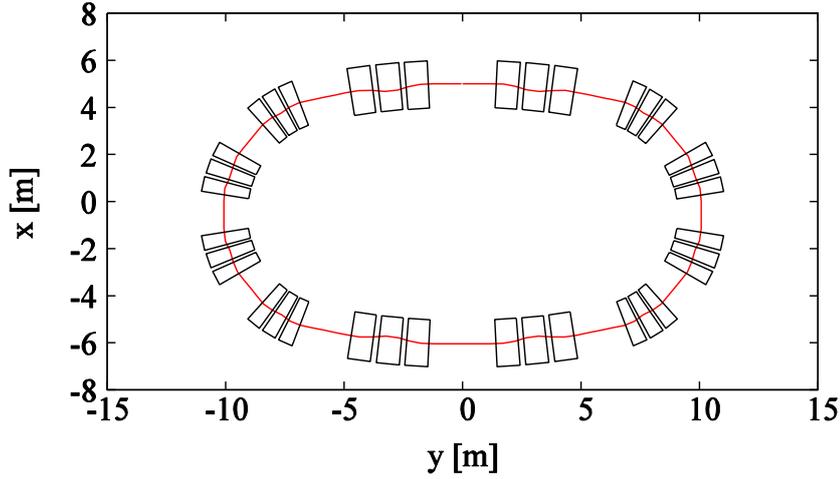

Figure 6: Layout of the advanced FFAG based on the arcs with different radii ("the egg-shaped ring").

The magnetic fields are linear in a non-scaling FFAG consisting only of dipole and quadrupole field components. In this way, the only non-linearities that could limit the machine acceptance come from fringe fields and kinematics. A machine with 10 cells was designed and encouraging tracking results were obtained. In order to reduce the orbit excursion, which is comparable to the scaling baseline design, a larger number of cells would be needed (~20). If symmetric cells with the long drift (2.2 m) would be used, the size of the ring would be twice the reference size and would correspond to h=2 operation within the frequency range of the MA (Magnetic Alloy) loaded RF cavities. This would require twice the RF voltage per turn in order to keep the momentum acceptance as in the initial design, which would be expensive. An alternative design principle [13] could be based on an arc consisting of the compact cells matched to the straight insertion with the long drifts with the help of the dispersion suppressor and the betatron matching section. Such a ring would be smaller in size, compatible with h=1 operation and have smaller orbit excursion. The preliminary study of the dispersion suppressor and the betatron matching section gives promising results. Fig. 6 shows half of the 180º arc with the dispersion suppressor at the beginning.

As the transverse-longitudinal coupling is present in ns-FFAGs due to a natural chromaticity, its effect on the final energy spread and beam quality needs to be tested. In order to gauge the expected results, an experiment was designed and will be performed on EMMA, if funding can be obtained.

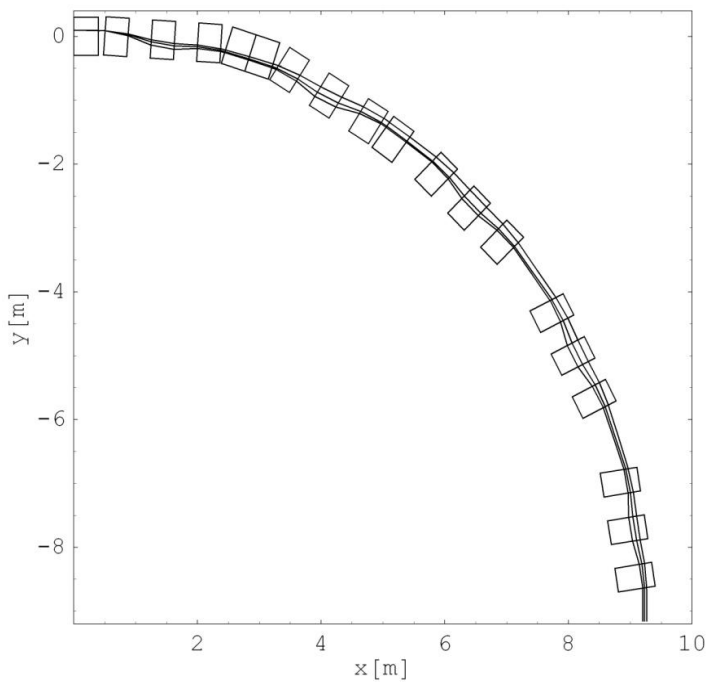

Figure 6: Half of the 180º non-scaling FFAG arc with the dispersion suppressor at the beginning followed by 3 regular triplet cells. Additional matching dipoles in the dispersion suppressor section are visible. The trajectories of reference particle and particles with ±20 % momentum deviation show a reasonable match.

### 4.3 Superperiodic FFAG

Another method to generate the long straight sections needed for injection and extraction is to introduce a superperiodicity. This can be done still keeping strictly to the scaling FFAG conditions. The field profile to satisfy the scaling conditions is

$$B_z = B_0 \left(\frac{r}{r_0}\right)^k F(\theta)$$

where $B_0$ is the vertical field at the reference radius $r_0$ and $F(\theta)$ describes the azimuthal dependence of the fields. The scaling FFAG design so far assumed a simple repetitive function for $F(\theta)$, such as a FODO or triplet. However, there is no reason why it cannot have a more complicated azimuthal dependence. In order to make enough space for injection, extraction and RF cavities, it is desirable to have a variety of drift space distances instead of many identical and rather short spaces.

As an example, a four fold symmetry FFAG lattice for PRISM was designed as shown in Fig. 7. Each arc consists of three triplet focusing units and an extra focusing magnet at both ends. The resulting number of magnets is rather large compared to the original scaling PRISM lattice, but it could be reduced by further optimization. The design also needs the adjustment of each magnet strength to eliminate beating of beta functions due to long drifts. A fitting procedure is established and implemented as a design code [14].

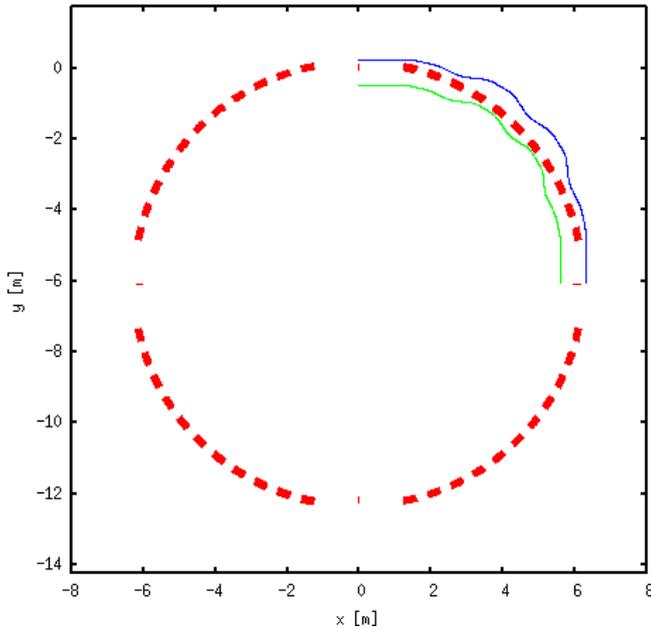

Figure 7: Scaling FFAG lattice with four fold symmetry. Red blocks indicate magnets. Green and blue lines are orbits whose momentum ratio is 2.6.

## 5. RF system developments

An RF system has been constructed and tested [15]. The beam experiment using alpha particle beam has been performed to simulate the bunch rotation [5]. Very large (~1.7 m X 1.0 m) magnetic alloy cores were loaded in the cavity. To drive the cavity, a compact and high peak power amplifier was developed for very low duty operation. Total RF voltage of 2-3 MV at 3.8 MHz is necessary and more than 40 RF systems are required. To reduce the cost, it is important to improve the performance of the magnetic alloy. To achieve that the work on development of a new material, FT3L, was launched [16], which led to a substantial progress in the design of MA cavities [17]. Large-size MA cores have been successfully fabricated using this new material at J-PARC. Those cores have two times higher impedance than ordinary FT3M MA cores. These developments may be used for the PRISM RF system in order to either reduce the core volume cutting the cost by a factor of 3 or to increase the field gradient. Both options should be considered.

## 6. Summary

High intensity and purity muon beams are needed for particle physics experiments in particular for the lepton flavour violation searches. FFAG rings are the best option to produce such beams in a cost effective way. The PRISM Task Force is working towards solving the remaining challenges on the path to this experiment and a substantial progress was achieved. The reference parameter set has been created and is shown in Table 3. The studies aim for the design of a feasible beam transport and injection/extraction system based on the existing technology. Several new ideas like the advanced FFAG or the superperiodic scaling FFAG have been proposed. The performance of the alternative designs will be evaluated and compared with the reference one. The design of hardware components like kicker magnets and RF cavities will be continued. The new concepts and ideas developed for PRISM may be applied for nuSTORM, the Neutrino Factory, the Muon Collider and also for cancer therapy, energy production and neutron sources.

Table 3. Accelerator parameters of the PRISM system for muon to electron conversion experiment.

| Proton beam parameters | |
|---|---|
| Beam power | 0.75-2 MW |
| Beam kinetic energy | 2-8 GeV |
| Bunch length at the pion production target | ~10 ns rms |
| Repetition rate | 1 kHz |
| **Target and pion/muon beam transport** | |
| Target type | solid |
| Capture element | solenoid 4-10 T |
| Transport system | solenoidal channel / FFAG transport line |
| Beam polarity | negative |
| **PRISM ring parameters** | |
| Machine function | Muon beam phase rotation and purification |
| Machine type | FFAG |
| Momentum acceptance | ±20 % |
| Reference muon momentum | 40-68 MeV/c |
| Minimal physical acceptance (H/V) | (3.8/0.57) π cm rad |
| Harmonic number | 1 |
| RF voltage per turn | 2-3 MV |
| RF frequency | 3-6 MHz |
| Injection/extraction type | single turn |
| Extraction kicker rise time | 50-60 ns |
| Repetition rate | 1 kHz |
| Initial beam momentum spread | ±20 % |
| Final beam momentum spread | ±2 % |
| Number of turns | ~6 |
| Number of synchrotron oscillations | 1/4 or 3/4 |